\documentclass[11pt,oneside,letterpaper]{article}
\usepackage{amssymb}
\usepackage{amsmath}
\usepackage[dvips]{graphicx}
\usepackage{setspace}
\usepackage{amssymb}
\usepackage{fancyhdr}
\usepackage{xcolor}
\usepackage{ifpdf}
\usepackage{graphicx}
\usepackage{rotating}
\usepackage{comment}
\usepackage{color}
\usepackage{cite}
\definecolor{darkgreen}{rgb}{0,0.5,0}
\definecolor{darkblue}{rgb}{0,0,0.6}
\definecolor{purple}{rgb}{0.4,.2,0.7}
\newcommand{\p}{\partial}
\newcommand{\h}{\theta}
\newcommand{\f}{\frac}
\newcommand{\be}{\begin{equation}}
\newcommand{\ee}{\end{equation}}
\usepackage[colorlinks=true,citecolor=darkgreen,linkcolor=purple,urlcolor=purple]{hyperref}

\numberwithin{equation}{section}
\allowdisplaybreaks  

\begin{document}
\begin{titlepage}

\vspace*{2.3cm}

\begin{center}
{ \LARGE \textsc{Late-time Structure of the Bunch-Davies \\ FRW Wavefunction}\\}
\vspace*{1.7cm}
George Konstantinidis$^1$, Raghu Mahajan$^1$ and Edgar Shaghoulian$^{2}$

\vspace*{0.6cm}
$^1$ {\it Stanford Institute of Theoretical Physics, Stanford University} \\
$^2$ {\it Department of Physics, University of California Santa Barbara}

\vspace*{0.6cm}

cgcoss@stanford.edu, rm89@stanford.edu, edgars@physics.ucsb.edu


\end{center}
\vspace*{1.5cm}
\begin{abstract}
\noindent

\noindent In this short note we organize a perturbation theory for the Bunch-Davies wavefunction in flat, accelerating cosmologies. The calculational technique avoids the in-in formalism and instead uses an analytic continuation from Euclidean signature. We will consider both massless and conformally coupled self-interacting scalars. These calculations explicitly illustrate two facts. The first is that IR divergences get sharper as the acceleration slows. The second is that UV-divergent contact terms in the Euclidean computation can contribute to the absolute value of the wavefunction in Lorentzian signature. Here UV divergent refers to terms involving inverse powers of the radial cutoff in the Euclidean computation. In Lorentzian signature such terms encode physical time dependence of the wavefunction. 

\end{abstract}
\end{titlepage}

\newpage
\setcounter{page}{1}
\pagenumbering{arabic}

\setcounter{tocdepth}{2}

\onehalfspacing

\clearpage

\section{Introduction}
Infrared divergences have a long and controversial history in the context of cosmology. Inflationary calculations often utilize the ``in-in" formalism of Schwinger and Keldysh \cite{Schwinger:1960qe, Keldysh:1964ud, Mahanthappa:1962ex, Bakshi:1962dv, Bakshi:1963bn, Calzetta:1986ey, Morikawa:1994jh, Tsamis:1994ca, Tsamis:1996qm, Tsamis:1997za, Onemli:2004mb, Prokopec:2002uw, Brunier:2004sb, Prokopec:2003qd, Collins:2005nu, Boyanovsky:2005sh, Senatore:2009cf, Senatore:2012nq, Pimentel:2012tw, Giddings:2010nc}.\footnote{For a review of the in-in formalism,  see \cite{Weinberg:2005vy, Adshead:2009cb}, and for a review of infrared effects in cosmology, see \cite{Seery:2010kh}.} A recent proposal for analyzing infrared issues in the presence of interaction is to deal directly with the  wavefunction for the fields of interest \cite{Anninos:2014lwa}. 
The technique of computing wavefunctions in cosmological spacetimes via analytic continuation from Euclidean signature was introduced earlier in \cite{Maldacena:2002vr}.
Computing the wavefunction directly helps us understand how divergences in the wavefunction will feed into divergences of any type of observable. The simplest analogy for this point of view involves quantum mechanics: the hydrogen atom is a well-defined quantum system even though observables like $\langle 1/r^3 \rangle$ diverge. The wavefunction point of view makes clear precisely which observables will diverge and which ones are well-defined. A more relevant example is that of a massless scalar field in de Sitter space (in any dimension). The equal-time correlator $\langle \phi(x) \phi(y) \rangle \sim \log( \vert x-y \vert \Lambda)$ has a long-distance divergence which is regulated by some IR cutoff $\Lambda$. The resolution to this divergence is to discard this as an unphysical (i.e. unobservable) correlation function. One should instead deal with observables like $\langle (\partial \phi)(x) (\partial \phi)(y)\rangle\sim \frac{1}{(x-y)^2}$ which tame the IR behavior.

An interesting extension of the perturbative corrections to the Bunch-Davies de Sitter wavefunction is to the one-parameter family of accelerating  Bunch-Davies FRW wavefunctions. This is the extension we will explore in this note. Equal-time correlators of free scalar fields in the Bunch-Davies vacuum for accelerating FRW have IR divergences which increase as the acceleration slows. This is an infrared spectral tilt. In particular, de Sitter has the softest IR structure, with only a logarithmic divergence and scale-invariant spectrum. We will particularly be interested in the pattern of spatial and temporal IR divergences as a function of the acceleration upon including interactions. This may help give some insight into which values of the acceleration parameter lead to IR divergences which can be consistently understood in terms of some putative Q-space/QFT duality \cite{Shaghoulian:2013qia}. (See also \cite{McFadden:2009fg, McFadden:2010na, McFadden:2010jw, McFadden:2010vh }.)

In \cite{Anninos:2014lwa}, perturbative corrections to the Bunch-Davies de Sitter wavefunction of self-interacting fields were found by calculating Witten diagrams in a fixed Euclidean AdS background and then performing an analytic continuation. We plan to follow a similar approach. As  noted in \cite{Shaghoulian:2013qia}, the FRW metric may be obtained by analytically continuing the metric of a Euclidean hyperscaling-violating geometry. Therefore, by computing Witten diagrams in a Euclidean hyperscaling-violating geometry and analytically continuing, we will have computed perturbative corrections to the Bunch-Davies wavefunction of self-interacting fields in a fixed FRW background.

We will consider two cases in our calculations. The first example will be a massless scalar with $\lambda\phi^4$ self-interaction. We will compute the perturbative corrections to the wavefunction at first order in $\lambda$, which constitutes evaluating tree-level and one-loop Witten diagrams. The second example will be a conformally coupled scalar field with general $\lambda\phi^n$ interaction. We will again compute the perturbative corrections to the wavefunction at first order in $\lambda$, which in this case constitutes evaluating all $L$-loop diagrams for $L=0,1,\dots, \lceil{n/2}\rceil-1$. 

In the rest of the introduction we will introduce and discuss the relevant geometries. In section \ref{massless} we will carefully build the diagrammatic expansion for the Bunch-Davies wavefunction for massless  scalar fields with $\lambda \phi^4$ self-interaction and compute the tree-level and one-loop corrections to the wavefunction. In section \ref{conformal} we will compute the tree-level and $L$-loop corrections to the wavefunction for a conformally coupled scalar with general $\lambda \phi^n$ interaction. We will stick to $(3+1)$ dimensions throughout, while appendix \ref{app:generald} will treat the conformally coupled scalar in arbitrary dimension. Appendix \ref{app:confmass} will present a non-minimally coupled scalar field theory with soluble wave equation that can also be analyzed. 

The flat FRW geometries are given by
\begin{equation}
ds_{d+1}^2=\ell^2 (-\eta)^{2\theta/(d-1)} \left(\frac{-d\eta^2+dx_i^2}{\eta^2}\right), 
\quad i \in \{1, \ldots d\}\,,\quad \eta<0\,,
\label{frwMetric}
\end{equation}
while the Euclidean hyperscaling-violating geometries \cite{Charmousis:2010zz, Huijse:2011ef} are given by 
\begin{equation}
ds_{d+1}^2=\ell^2 z^{2\theta/(d-1)}\left(\frac{dz^2+dx_i^2}{z^2}\right).
\quad i \in \{1, \ldots d\}\,,\quad z>0\,.
\label{hsvMetric}
\end{equation}
The analytic continuation from Euclidean signature \eqref{hsvMetric} to Lorentzian signature \eqref{frwMetric} is achieved via the following transformation:
\begin{align}
z &\to (-i) (-\eta),\label{eq:zcontinuation} \\
\ell &\to (-i)^{1-\theta/(d-1)}\ell.\label{eq:lcontinuation}
\end{align}
 When $\theta=0$, this is the usual connection between the de Sitter and Euclidean anti-de Sitter spacetimes. We will restrict ourselves to $\h \leq 0$, which correspond to accelerating cosmologies and are regular as $z\rightarrow 0$. 
 
A simple matter action which sources this background is a minimally coupled scalar with self-interacting potential:
\be 
S=\f{1}{2} \int d^{d+1}x \,\sqrt{g}\left(R-2(\p \Phi)^2 - \f{V_0^2}{\ell^2} e^{-\beta \Phi}\right)\,,
\ee 
\be 
V_0^2 = \f{1}{4}(d-1)(2-\h)(d(2+\h)-\h)\,,\qquad \beta = 2 \sqrt{\f{2\h}{(d-1)(\h-2)}}\,.
\ee 
The scalar field takes the on-shell value 
\be 
\Phi = \f{1}{2} \sqrt{\f{\h(d-1)(\h-2)}{2}} \log (-\eta)\,.
\ee 
For top-down constructions of such models, see e.g. \cite{Hellerman:2006nx, Dodelson:2013iba}.

In the context of inflationary physics, such models fall under the class of ``power-law inflation" \cite{Lucchin:1984yf}. The power law refers to the scale factor in synchronous coordinates $a(t) \sim t^q$ with $q>1$. To linear order in the slow-roll parameters $\epsilon_{sr}$ and $\eta_{sr}$ we have 
\be 
\epsilon_{sr} \sim -\h, \qquad \eta_{sr} \sim -2\h\,.
\ee 
The simplest versions of these models are disfavored by recent data \cite{Planck:2013jfk} (tuning $\h$ to accommodate for the infrared spectral tilt leads to a tensor mode contribution which is too large), but they are useful due to their analytic solubility. The nonlinear stability of this system is shown in \cite{Ringstrom:2009zz}. The particular matter theory which sources the FRW background will not, however, be important to us.

\section{Massless self-interacting scalar}\label{massless}
Consider a massless scalar field $\phi$ 
on the hyperscaling-violating background \eqref{hsvMetric}. 
This scalar is distinct from the scalar $\Phi$ from the previous section. 
We take the action to consist of the kinetic term and a $\phi^4$ interaction which is treated perturbatively:
\begin{align}
S[\phi] = \int d^{d+1}x\, \sqrt{g}\, \left( - \frac{1}{2}g^{\mu\nu} 
\partial_\mu \phi \partial_\nu \phi - \frac{\lambda}{24}\phi^4
\right).
\end{align}
The equation of motion of the free theory is
\begin{equation}
\partial_z \left(\sqrt{g}\, g^{zz}\partial_z{\phi(z)}\right)-\sqrt{g}\,g^{zz}k^2\phi(z)=0,
\label{masslessScalarEOM}
\end{equation}
where we are working in momentum space in the $x^i$ directions.
The general solution of (\ref{masslessScalarEOM}) is given by
\begin{align}
\phi(z)&=c_1 \phi_1(z) +c_2 \phi_2(z),
\quad \text{where} \\
\phi_1(z)=z ^{\frac{d-\theta }{2}}~&I_{\frac{d-\theta }{2}}\left(k z \right),\qquad
\phi_2(z)=z ^{\frac{d-\theta }{2}}~K_{\f{d-\theta }{2}}\left(k z \right).
\end{align}
Here $I$ and $K$ are Bessel functions.
To find the tree-level wavefunction, we need to find the on-shell action of the 
solution that is regular as 
$z\to \infty$ and satisfies Dirichlet boundary conditions at the cutoff surface 
$z = z_c \ll 1$. The Dirichlet boundary condition reads
\begin{equation}
\lim_{z \to z_c} \phi(z; \vec{k}) = \phi_{\vec{k}}\,,
\end{equation}
where $\phi_{\vec{k}}$ is some specified field configuration.
The solution with this property is easily found:
\begin{align}
\phi(z;\vec{k}) = \phi_{\vec{k}}\,  
\frac{z ^{\frac{d-\theta }{2}}~K_{\f{d-\theta }{2}}\left(k z \right)}
{z_c ^{\frac{d-\theta }{2}}~K_{\f{d-\theta }{2}}\left(k z_c \right)}\,.
\label{eq:phizkphikrelation}
\end{align}
Next, we compute the on-shell action, which is a functional of the boundary value of the field
\begin{align}
S_{\text{on-shell}}[\phi] &= \int d^dx\, \sqrt{g}\, g^{zz} \phi \partial_z \phi 
\Big \vert_{z = z_c}\\
&=\int\! \frac{d^dk}{(2\pi)^d}\, z_c^{\theta+1-d} \phi_{-\vec{k}}~
\phi_{\vec{k}} \left(
- k \frac{K_{\frac{2-d+\theta}{2}}(kz_c)}{K_{\frac{d-\theta}{2}}(kz_c)}
\right).
\end{align}
The wavefunction is then determined as
\begin{align}
\Psi_0[\phi] = \mathcal{N}\exp \left[ - S_{\text{on-shell}}
\right].
\end{align}
Expanding this out for small $z_c$, we see that the piece which 
will give us the cosmological correlator is
\begin{align}
\log \Psi_0 \sim - \int\! \frac{d^dk}{(2\pi)^d}\, \phi_{-\vec{k}}\phi_{\vec{k}}\, k^{d-\theta}.
\end{align}
After analytic continuation via \eqref{eq:zcontinuation}  and \eqref{eq:lcontinuation}, this wavefunction satisfies the Schr\"odinger equation for the Hamiltonian of a free scalar field in the FRW background:
\begin{align}
H = \frac{1}{2}\int \frac{d^dk}{(2\pi)^d} \left( 
-\frac{\vert \eta \vert ^{d-1-\theta}}{\ell^{d-1}}
 \frac{\delta}{\delta \phi_{-\vec{k}}}
 \frac{\delta}{\delta \phi_{\vec{k}}} +
\frac{\ell^{d-1}}{\vert \eta \vert ^{d-1-\theta}} k^2 \phi_k \phi_{-k}
\right).
\end{align}

\begin{figure}[t]
  \centering
    \includegraphics[width=0.8\textwidth]{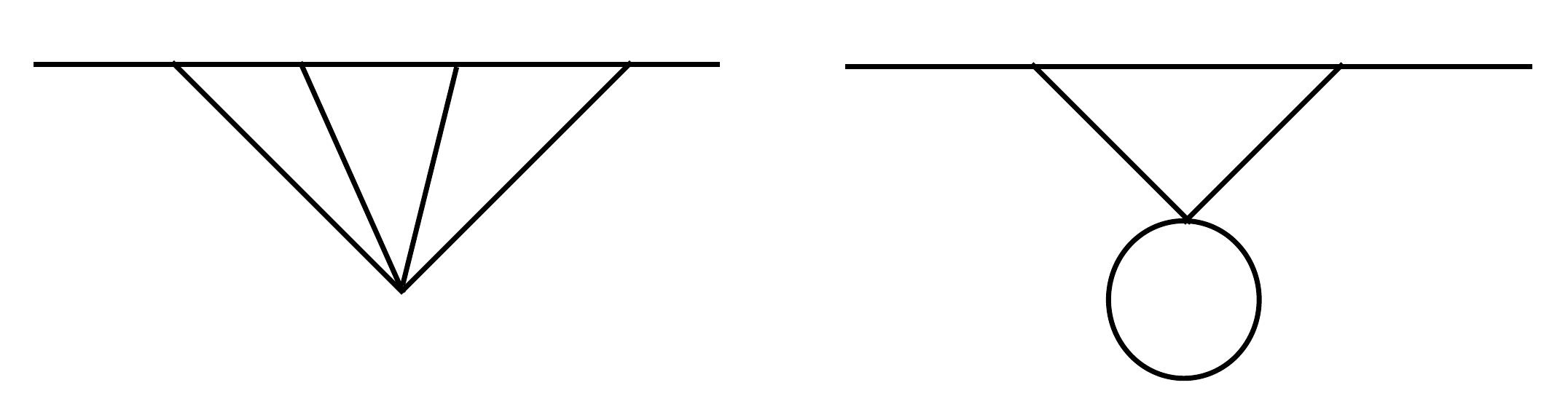}
    \caption{Witten diagrams for the order $\lambda$ contributions to the partition function in the Euclidean hyperscaling violating geometry, with $\phi^4$ interaction. These corrections analytically continue to corrections of the wavefunction in FRW spacetimes.}\label{phi4wittendiag}
\end{figure}

We now wish to incorporate self-interactions of the scalar and compute interaction corrections to the free wavefunction obtained above. Formally, to order $\lambda$, we have
\begin{align}
\log \Psi[\phi] &= \log \Psi_0[\phi]  \, + 
\delta\Psi_\lambda^{(4)}[\phi]  + \delta\Psi_\lambda^{(2)}[\phi]  +\ldots\,,\\
\delta\Psi_\lambda^{(4)} &= - \lambda \int \prod_{i=1}^4 \frac{d^dk_i}{(2\pi)^d} 
\mathcal{M}_\lambda^{(4)} (\vec{k}_1, \vec{k}_2, \vec{k}_3, \vec{k}_4)
\phi_{\vec{k}_1}\phi_{\vec{k}_2}\phi_{\vec{k}_3}\phi_{\vec{k}_4}\,,\\
\delta\Psi_\lambda^{(2)} &= - \lambda \int \prod_{i=1}^2 \frac{d^dk_i}{(2\pi)^d} 
\mathcal{M}_\lambda^{(2)}(\vec{k}_1, \vec{k}_2) \phi_{\vec{k}_1}\phi_{\vec{k}_2}.
\end{align}
The two corrections, $\mathcal{M}_\lambda^{(4)}$
and $\mathcal{M}_\lambda^{(2)}$, come from the two Witten diagrams in figure \ref{phi4wittendiag}.
In order to compute these Witten diagrams, we need to derive the bulk-to-bulk and bulk-to-boundary propagators for 
the massless scalar field on the background \eqref{hsvMetric}. 
These propagators
are then used to compute Witten diagrams.

\subsection{Propagators}
To construct the bulk-to-bulk propagator we need to identify the modes that diverge to plus and minus infinity as $z\rightarrow\infty$ \cite{Anninos:2014lwa}. These modes are just $\phi_1$ and $\phi_2$ respectively. 
The form of the bulk-to-bulk propagator is the following:
\begin{equation}
G(z,w;k)= 
\begin{cases} 
A\phi_1(z)\phi_2(w)+C\phi_2(z)\phi_2(w)\quad\mathrm{for}~w>z,\\
B\phi_2(z)\phi_1(w)+C\phi_2(z)\phi_2(w)\quad\mathrm{for}~z>w.
\end{cases}
\label{GzgwGeneral}
\end{equation}
We will now impose various conditions on the propagator given above in order to determine the unspecified coefficients. First, we require that $G(z,w,k)$ is continuous at $z=w$. This condition imposes that $A=B$. The second condition requires that as $z$ increases through $z=w$, the first derivative $\partial_zG(z,w,k)$ decreases by $z^{d-\theta-1}$. This condition implies that $A=-\frac{i \pi}{4}\,\ell^{1-d}$. Finally, we enforce the Dirichlet boundary condition $G(z=z_c,w,k)=0$ at the cutoff, which specifies $C$. The bulk-to-bulk propagator for $w>z$ can finally be written as:
\begin{equation}
G(z,w;k)=\ell^{1-d}w^{\frac{d-\theta }{2}} z^{\frac{d-\theta }{2}}\frac{ K_{\frac{d-\theta }{2}}\left(k w\right) }{K_{\frac{d-\theta }{2}}\left(k z_c\right)}\left(K_{\frac{d-\theta }{2}}\left(k z_c\right) I_{\frac{d-\theta }{2}}\left( k z\right)-K_{\frac{d-\theta }{2}}\left(k z\right) I_{\frac{d-\theta }{2}}\left(k z_c\right)\right).
\label{GwgzFinal}
\end{equation}
To get $G(z,w;k)$ for $w < z$, we simply swap $z$ and $w$ in the above equation.

We can obtain the bulk-to-boundary propagator via its relation to the bulk-to-bulk propagator:
\begin{equation}
K(z;k)=\sqrt{g(w)}\,g^{ww}\partial_w G(w,z; k)\bigg|_{w=z_c}.
\label{btbRelation}
\end{equation}
Plugging in for the expression obtained previously for $G(w,\vec y;z,\vec x)$ we find that
\begin{equation}
K(z; k)=\frac{z^{\frac{d-\theta }{2}}  K_{\frac{d-\theta }{2}}\left(k z\right)}{z_c^{\frac{d-\theta}{2}}K_{\frac{d-\theta }{2}}\left(k z_c\right)},
\label{btboundFinal}
\end{equation}
an expression which could also have been obtained directly from \eqref{eq:phizkphikrelation}.

\subsection{Tree-level diagram for the four-point function}
From now on we will work in $d=3$ corresponding to a four spacetime dimensions.
The tree level diagram (drawn on the left in Figure \ref{phi4wittendiag}) is given by 
\begin{align}
\mathcal{M}_{\lambda}^{(4)}(k_1, k_2, k_3, k_4)=
\int dz \sqrt{g}\, K (z;k_1) K(z;k_2) K(z;k_3) K(z;k_4). 
\end{align}
The small $z$ behavior of the bulk-to-boundary propagators is smooth and nonvanishing, and the
singularities all come from the $\sqrt{g}$ factor:
\begin{align}
\int dz \sqrt{g} \sim \int dz\, z^{2\theta-4} \sim z^{2\theta-3}\,.
\end{align}
Doing the integrals for a generic value of $\theta$ and expanding for 
small $z_c$, we see that there are two power series in the expression for the tree level diagram. Schematically, we can write the two power series as follows: 
\begin{align}
\mathcal{M}_{\lambda}^{(4)}(k_1, k_2, k_3, k_4)  \sim \ell^4 \big( &z_c^{2\theta-3} + k^2 z_c^{2\theta-1} + k^4 z_c^{2\theta+1} + \ldots \label{firstmasslesstree}\\
&+ k^{3-\theta}z_c^{\theta} + k^{5-\theta}z_c^{\theta+2} + k^{7-\theta}z_c^{\theta+4} + \ldots \big)\label{secondmasslesstree},
\end{align}
where we have not written down the coefficients in front of any of the terms. We also use the generic notation $k^\alpha$ to refer to a homogeneous order $\alpha$ polynomial in the four variables $k_1, \ldots k_4$.
Let us look at some special values of $\theta$.

When $\theta$ is a negative odd integer, $z_c^\theta$ appears in the first power series. 
In this case, the second power series starts with $z_c^\theta \log z_c$.
To be concrete, for $\theta = -1$, we have\
\begin{align}
\mathcal{M}_{\lambda}^{(4)}(k_1, k_2, k_3, k_4)  &= \ell^4 \left(
- \frac{1}{5z_c^5} + \frac{\sum_i k_i^2}{30z_c^3} + \ldots
\right) + 
\ell^4 \left( 
\frac{\sum_i k_i^4 \log(k_i z_c/2)}{20 z_c}
+ \ldots \right).
\end{align}

When $\theta$ is zero or a negative even integer, the second power series contains a $\log z_c$ term. 
The $\theta=0$ case considered in \cite{Anninos:2014lwa} falls under this category where the second power series starts with $\log z_c$. To be concrete, for $\theta = -2$, we have
\begin{align}
\mathcal{M}_{\lambda}^{(4)}(k_1, k_2, k_3, k_4)  &= \ell^4 \left(
- \frac{1}{7 z_c^7} + \frac{\sum_i k_i^2}{105 z_c^5} + \ldots
\right)
+ \ell^4 \left( 
-\frac{\sum_i k_i^5 }{126 z_c^2} 
+ c_1 \log z_c \ldots \right).
\label{eq:thetaminus2}
\end{align}
The constant $c_1$ depends on the momenta. The expression is quite ugly, and therefore we have not written it down explicitly.

Our job now is to continue these Euclidean-signature calculations to our Lorentzian FRW metric. Let us emphasize that we do not subtract off the terms with inverse powers of $z_c$ via holographic renormalization. This is because in the Lorentzian signature, they encode physical time-dependence of the wavefunction. 

The analytic continuation to Lorentzian signature is achieved via the transformation in equations \eqref{eq:zcontinuation} and \eqref{eq:lcontinuation}.
In analytically continuing the first power series, the factor of $(-i)^{-2\theta}$ coming from the continuation of $\ell^4$ combines with the $(-i)^{2\theta-3}$ from the first term to give $(-i)^{-3}$, which is independent of $\theta$ and purely imaginary. The step size is by two and so every term in the series is pure imaginary.
In the second power series, the imaginary factors coming 
from the analytic continuation depend on $\theta$. We get
$(-i)^{-2\theta} \times (-i)^{\theta} = (-i)^{-\theta}$. Since $(-i)^{-\theta}$ has a real part for generic $\theta$, this means that we get a contribution to $\vert \Psi \vert$. 

In particular, for $\theta=-2$, the contribution from the second series in \eqref{eq:thetaminus2} is purely real and thus contributes to $\vert \Psi \vert$. Let us focus on the first term in the second series in (\ref{eq:thetaminus2}). This term in UV-divergent since it has inverse powers of $z_c$. It is also a contact term in position space, which we can see by inverse Fourier transforming $\sum_i k_i^5$ to position space. 

Thus, we see several situations where a local UV-divergent piece in the Euclidean calculation analytically continues to the magnitude of the wavefunction instead of its phase. This is one of the main messages of this paper.

\subsection{One-loop diagram for the two-point function}
In the $\lambda \phi^4$ theory at order $\lambda$, we also have a one-loop diagram, drawn on the right in Figure~\ref{phi4wittendiag}. It evaluates to
\begin{align}
\int_{z_c}^\infty dz\, \sqrt{g}\, K(z; \vec{k})K(z; \vec{k}) 
\int \frac{d^3p}{(2\pi)^3}\, G(z,z;p).
\end{align}
We first have to integrate $G$ over the loop momentum $p$.
A physical UV cutoff is imposed on the loop momentum $p$ as
\begin{align}
p_{\text{cutoff}} = \frac{\Lambda}{z^{1-\theta/(d-1)}}\, ,
\end{align}
where the power of $z$ is determined by the metric.
Power counting establishes that this term is quadratically divergent
\begin{align}
\int dp\,p^2\, G(z,z; p) \sim z^{-\theta} \Lambda^2.
\end{align}
The two power series we get in this case for generic $\theta$ are
\begin{align}
\mathcal{M}_\lambda^{(2)}(\vec{k}) = \,
\ell^2 \Lambda^2 \Big(&z_c^{\theta-3} + k^2 z_c^{\theta-1} + k^4 z_c^{\theta+1} + \ldots \nonumber \\
&+  k^{3-\theta}\log z_c + k^{5-\theta}z_c^2\log z_c+ \ldots \Big),\label{generalloop}
\end{align}
where again we have just exhibited the general structure and not written out the coefficients in front of any of the terms.
The only $z_c$-divergent term in the second series is the first one.
The special case to be noted
is that when $\theta$ is a negative odd integer, $\log z_c$ appears in the first power series: In this case, the second power series starts with $(\log z_c)^2$. 

The power of $\ell$ appearing in $\mathcal{M}_\lambda^{(2)}$ 
analytically continues as $\ell^2 \to \ell^2 \times (-i)^{-\theta}$.
The factor of $(-i)^{-\theta}$ coming from the continuation of $\ell^2$ combines with the $(-i)^{\theta-3}$ from the first power series to give $(-i)^{-3}$. The $\theta$ dependence in the analytic continuation cancels in the first series, as what happened for the tree-level case. The second series, in general, contributes divergent terms to the magnitude of the wavefunction.

Note that while we have exhibited the divergent structure of the loop diagram, it must be borne in mind that it is a tadpole diagram. Being a tadpole diagram, it can be exactly canceled by a counterterm in the original scalar action of the form
\begin{align*}
c\lambda\int d^4x \sqrt{g}\,  \phi^2,
\end{align*} 
where we can tune the coefficient $c$ sitting in front of this term.

\section{Conformally coupled scalar}\label{conformal}
The conformally coupled scalar is a useful test case to consider due to its analytic solubility. The mode functions are directly related to those of de Sitter due to the conformal equivalence of the two backgrounds. The action for our conformally coupled scalar is given as 
\begin{align}
S= \int d^{d+1} x \sqrt{g}\left(- \frac{1}{2}g^{\mu\nu}\partial_{\mu}\phi\partial_{\nu}\phi+\frac{d-1}{8d}R\phi^2-\frac{\lambda_n}{n!}\phi^n\right)
\end{align}
and has the invariance $g_{\mu\nu} \rightarrow \Omega^2 g_{\mu\nu}$, $\phi \rightarrow \Omega^{(1-d)/2} \phi$. Picking $\Omega = z^{\theta/(d-1)}$ maps anti-de Sitter to a hyperscaling-violating spacetime, and the mode functions will therefore transform as $\phi \rightarrow z^{-\theta/2} \phi$. The wave equation of the free theory is given as 
\begin{align}
\left(\f{1}{\sqrt{g}}\p_\mu \left(\sqrt{g}g^{\mu\nu}\p_\nu\right) + \frac{d-1}{4d} R\right) \phi = 0,\qquad R=-\f{d\left((d-\h)^2-1\right)}{\ell^2(d-1)}\;z^{-\f{2\h}{d-1}}
\end{align}
with solution
\begin{align}
\phi(z)=c_1 \phi_1(z)+c_2 \phi_2(z)\,;\qquad \phi_1(z)=z^{\f{-1+d-\h}{2}} \,e^{kz}, \quad \phi_2(z)=z^{\f{-1+d-\h}{2}} \f{e^{-kz}}{2k}\,.
\end{align}
Notice that the $\h=0$ case is related to the $\h\neq 0$ case as advertised. 

The bulk-to-bulk propagator is given as 
\begin{align}
G(z,w,k)&=A\phi_1(z)\phi_2(w)+C\phi_2(z)\phi_2(w)~~\mathrm{for}~w>z,\label{GwgzGeneralConf}\\
G(z,w,k)&=B\phi_2(z)\phi_1(w)+C\phi_2(z)\phi_2(w)~~\mathrm{for}~z>w,\label{GzgwGeneralConf}\\
B&=A=\ell^{1-d}, \qquad C=-2k \,e^{2kz_c} \,\ell^{1-d}.
\end{align}
Altogether this gives the following bulk-to-bulk and bulk-to-boundary propagators:
\begin{align}
G(z,w,k)&=\frac{\ell ^{1-d}}{2k} e^{-k (w+z)} \left(e^{2 k z}-e^{2 k z_c}\right) (w z)^{\frac{1}{2} (d-\theta-1)}\,, \\
 K(z,z_c)&= \left(\f{z}{z_c}\right)^{\frac{1}{2}
   (d-\h-1)} e^{-k (z-z_c)}\,.
\end{align}
\subsection{Tree-level diagram}
Let us consider an arbitrary interaction $\phi^n$ for integer $n\geq 3$ in $d=3$. The case of general $d\neq 3$ will be treated in Appendix \ref{app:generald}. The tree-level diagram (drawn on the left in Figure~\ref{phi4wittendiag} but now with $n$ legs going out to the boundary) evaluates to
\begin{align}
\mathcal{M}^{(n)}_\lambda = \int dz \sqrt{g} \prod_{i=1}^n &K(z,k_i)=\ell^4 e^{k_{\Sigma_n} z_c} z_c^{2\theta-3} E_{\left(-\f{(n-4)(2-\theta)}{2}\right)}(k_{\Sigma_n} z_c)\,,\quad k_{\Sigma_n} = \sum_{i=1}^n k_i\,.
\end{align}
for the exponential integral function $E_{(\nu)}(z) = \int_1^\infty dt e^{-zt}/t^\nu$. Expanding this general expression for small $z_c$, the divergent structure for $n \neq 3$ contains no logarithms and is as follows:
\begin{align}
&\mathcal{M}_{\lambda}^{(n)}=\,\ell^4
\left(z_c^{2\theta-3}+k_{\Sigma_n}z_c^{2\theta-2}+\dots\right.\label{firstconformaltree}\\
+&\left.k_{\Sigma_n}^{3-2\theta +\f{n}{2}(\theta-2)}z_c^{\f{n}{2}(\theta-2)}+k_{\Sigma_n}^{4-2\theta +\f{n}{2}(\theta-2)}z_c^{1+\f{n}{2}(\theta-2)}+\dots\right).\label{secondconformaltree}
\end{align}
For even $n$ and negative-odd-integer $\theta$, or for odd $n\neq 3$ and vanishing or negative-even-integer $\theta$, the two series truncate to a finite number of terms:
\be \mathcal{M}_{\lambda}^{(n)}=\ell^4\left(k^{3-2\theta+\f{n}{2}(\theta-2)}z_c^{\f{n}{2}(\theta-2)} +k^{4-2\theta+\f{n}{2}(\theta-2)}z_c^{1+\f{n}{2}(\h-2)}+\cdots+z_c^{2\theta-4}/k\right).
\ee
The case of $n=4$ for any $\h$ is particularly simple:
\footnote{The expression takes this simple form because a quartic interaction in four dimensions preserves conformal invariance of the scalar field. Indeed, we can formally consider the classically scale-invariant interaction $\phi^{2(d+1)/(d-1)}$ in arbitrary dimension (formal since this is non-integer unless $d=3, 5, \infty$). The entire answer in this case is
\be
\mathcal{M}_{\lambda}^{(2(d+1)/(d-1))}=-\ell^{d+1}z_c^{\f{(d+1)(\theta-d+1)}{d-1}}/k_\Sigma\,.
\ee
The numerical coefficient, which usually we ignore in our schematic expansions but restore here, is simply $-1$. Its independence of $\theta$ is related to the fact that $n=4$ preserves conformality at tree level, and  $\theta \neq 0$ is conformally related to $\theta = 0$.
}
\be
\mathcal{M}_{\lambda}^{(4)}=-\ell^4z_c^{2\h-4}/k_\Sigma\,.
\ee
 For the case of $n=3$ we have logarithms appearing for negative-even-integer or vanishing $\theta$:
\begin{align}
\f{\mathcal{M}^{(3)}_{\lambda, \,\theta=0}}{\ell^4} = &
-\f{\gamma_E+\log(k_\Sigma z_c)}{z_c^3}-
\f{k_\Sigma}{z_c^2}\left(-1+\gamma_E+\log(k_\Sigma z_c)\right)\nonumber\\
&-\f{k_\Sigma^2}{4z_c}\left(-3+2\gamma_E+2\log(k_\Sigma z_c)\right)
-\f{k_\Sigma^3}{36}(-11+6\gamma_E+6\log(k_\Sigma z_c))+\dots\\
\f{\mathcal{M}^{(3)}_{\lambda, \,\theta=-2}}{\ell^4} &= \f{1}{z_c^7}+
\f{k_\Sigma}{z_c^6}(\gamma_E+\log(k_\Sigma z_c))
-\f{k_\Sigma^2}{z_c^5}(-1+\gamma_E+\log(k_\Sigma z_c))\nonumber\\
-\f{k_\Sigma^3}{4z_c^4}&\left(-3+2\gamma_E+2\log(k_\Sigma z_c)\right)
-\f{k_\Sigma^4}{36z_c^3}(-11+6\gamma_E+6\log(k_\Sigma z_c))+\dots\,.
\end{align}
In this case, unlike the rest of the expressions in this note, we have explicitly displayed the numerical coefficients of the expansion. This is to illustrate the matching numerical coefficients appearing between the two series. The general structure for vanishing or negative-even-integer $\theta$ is
\begin{align}
\mathcal{M}^{(3)}_{\lambda, \,\theta} = \ell^4\sum_{i=1}^{\infty} \ z_c^{2\theta-3+i-1}k^{i-1}[c_i(\theta)+d_i(\theta)(\log(k_\Sigma z_c))^{m_i}]\,,\quad m_i = \textrm{max}\{0,i+\theta/2\}
\end{align}
for constants $c_i$, $d_i$ satisfying the interesting relations $c_i(\theta) =- c_{i-\theta/2}(\theta-2)$ and $d_i(\theta) =- d_{i-\theta/2}(\theta-2)$, which relate different theories.

The first power series \eqref{firstconformaltree} looks just like the first power series for the massless scalar with $\phi^4$ interaction \eqref{firstmasslesstree}, except in this case the step size is by single powers of $z_c$ instead of $z_c^2$. This means that the second, fourth, etc., terms in \eqref{firstconformaltree}, which are local in position space, analytically continue to the \emph{real} part of the wavefunction. Even for the renormalizable interaction $\phi^3$, for generic $\h$ these local terms exist and continue to the real part of the wavefunction, and for negative-even-integer or vanishing $\h$ the logarithms that appear contribute to the real part of the wavefunction as well. This contribution to the real part of the wavefunction from a local term in position space was noticed for $\h=0$ in \cite{Anninos:2014lwa}.

\subsection{One-loop diagram}
Let us now consider the one-loop diagram drawn on the right in Figure~\ref{phi4wittendiag}, but with $n-2$ legs going out to the boundary. We want to calculate
\begin{align}
\mathcal{M}_{\lambda}^{(n-2)} &= \int_{z_c}^{\infty} dz \sqrt{g} \prod_{i=1}^{n-2} K(z,k_i)\int_0^{\Lambda \ell/z^{1-\theta/2}} d^3 p \, G(z,z;p)\,.
\end{align}
We first do the momentum integral of the bulk-to-bulk propagator to obtain 
\be
\f{1}{2} \pi  z^{-\theta} \left(\frac{z e^{2 \Lambda  \ell  z^{(\theta-2)/2} (z_c-z)} }{\ell^2(z-z_c)^2} \left(2 \Lambda  \ell  z^{\theta/2} (z-z_c)-z e^{2 \Lambda  \ell  z^{\f{\theta}{2}-1} (z-z_c)}+z\right)+2 \Lambda ^2 z^{\theta}\right)\,.
\ee
There are different regularization procedures one can adopt at this point. We will keep the leading divergence $\Lambda^2$ and drop the rest, which is finite as $\Lambda \rightarrow \infty$. Although the finite piece can be kept and its contribution to the wavefunction calculated, it is subleading. Integrating the leading divergence against the $n-2$ bulk-to-boundary propagators gives us
\be
\mathcal{M}_{\lambda}^{(n-2)}=\pi  \Lambda ^2 \ell ^4 e^{k_{\Sigma_{n-2}} z_c} z_c^{2 \theta-3} E_{\left(\f{1}{2} (n-6) (\theta-2)\right)}(k_{\Sigma_{n-2}} z_c)\,.
\ee
This is the same expression as for the tree-level diagram, except with $n\rightarrow n-2$. This is easy to understand: the quadratic divergence $\Lambda^2$ coming from the momentum integral over the bulk-to-bulk propagator is independent of $z$ and $z_c$. Thus, the small $z_c$ expansion is the same as in the tree-level case, after accounting for the shift $n\rightarrow n-2$. As mentioned in the case of the massless scalar field, these loop diagrams are tadpole diagrams and can be canceled by counterterms in the action.

\subsection{$L$-loop diagram}
We can generalize the results of the previous subsection to arbitrary loop order $L\leq \lceil n/2 \rceil-1$, at first order in $\lambda$. Then we need to compute
\begin{align}
\mathcal{M}_{\lambda}^{(n-2L)} &= \int_{z_c}^{\infty} dz \sqrt{g} \left(\prod_{i=1}^{n-2L} K(z,k_i)\right)\int_0^{\Lambda \ell/z^{1-\theta/2}} \left(\prod_{i=1}^L d^3 p_i  G(z,z;p_i)\right)\\
&=(\pi  \Lambda ^2)^L \ell ^4 e^{k_{\Sigma_{n-2L}} z_c} z_c^{2 \theta-3} E_{\left(\f{(\theta-2) (n-2L-4)}{2}\right)}(k_{\Sigma_{n-2L}} z_c)\,.
\end{align}
This is the same as the tree-level answer, except with a divergent factor of $\Lambda^{2L}$ and $n\rightarrow n-2L$, as expected.

\section{Conclusions}
We have used the connection from \cite{Shaghoulian:2013qia} between Euclidean hyperscaling-violating spacetimes -- which appear generically within holography for non-conformal branes -- and accelerating FRW cosmologies to organize a perturbation theory for the Bunch-Davies wavefunction. This generalizes the procedure first discussed in \cite{Anninos:2014lwa}. 

One reason to focus directly on the cosmological wavefunction is the clarity of well-defined observables, as explained in the introduction. Another reason to focus on the cosmological wavefunction is that it plays a starring role in holographic duality. Some of the results of \cite{Anninos:2014lwa} where $\h=0$, for example the lack of logarithmic IR divergences for the spin-1 and spin-2 fields, can be understood and interpreted via the dS/CFT duality \cite{Maldacena:2002vr, Strominger:2001pn, Witten:2001kn}. In particular, since gauge fields are dual to conserved currents, and logarithmic IR divergences can be interpreted as shifts in operator dimensions, the lack of divergences is simply the statement that the currents remain conserved and their anomalous dimensions therefore vanish. As any purported field theory dual to  an FRW phase is not expected to be a CFT unless $\h=0$, there is no reason to expect these fields to be protected in a similar way. Certain perturbative wavefunctions, for example those of a conformally coupled scalar field or higher-spin fields, can potentially make contact with wavefunction calculations \cite{Anninos:2012ft, Anninos:2013rza, Conti:2014uda, Hartle:2007gi, Anninos:2012qw, Banerjee:2013mca} in non-minimal higher-spin dS/CFT \cite{Anninos:2011ui, Anninos:2014hia, Chang:2013afa}. 

Besides elaborating on our proposal for calculating correlators in accelerating FRW spacetimes with a few concrete toy models, there are two conceptual upshots to our calculations. The first concerns an often-repeated folk theorem that divergences which are contact terms in position space in Euclidean signature analytically continue to phases of the wavefunction. This connection was first highlighted in the context of tree-level diagrams in dS$_4$ \cite{Maldacena:2002vr}. The examples considered in our work explicitly illustrate that this is not generally true. This should come as no surprise, and violations of this folk theorem already appeared at the level of perturbative interactions in dS$_4$ \cite{Anninos:2014lwa}, but it seems to be a confusion that refuses to go away. 

The second upshot is that the late-time and large-space infrared divergences of the Bunch-Davies wavefunction due to self-interactions worsen as the acceleration slows. Concretely, this means that as $\theta$ becomes more and more negative, the leading power-law term in the small-$z_c$ expansion becomes more and more singular; see, for example, equations \eqref{firstmasslesstree} and \eqref{firstconformaltree}. 
This mirrors the fact that the large-space infrared divergences of the free theory worsen as the acceleration slows. This extreme infrared structure in the large-space correlation function has been shown to imply a fascinating ultrametric structure in the state space of a scalar field for $\h=0$ in \cite{Anninos:2011kh, Roberts:2012jw}. This ultrametric structure becomes sharper as the acceleration slows \cite{Shaghoulian:2013qia}, i.e. as $\h$ decreases, and the perturbative corrections indicate an analogous story. What this extreme infrared structure implies for a holographic dual theory remains an open question.

\section*{Acknowledgements} 
The authors would like to thank Dionysios Anninos, Tarek Anous, Daniel Green, Juan Maldacena, and Daniel Z. Freedman for stimulating discussions and comments. RM is supported by a Gerhard Casper Stanford Graduate Fellowship.

\appendix
\section{Conformally coupled scalar in general dimension}
\label{app:generald}

\subsection{Tree-level diagram}
Let us consider an arbitrary interaction $\phi^n$ for integer $n\geq 3$ in general dimension. The tree-level diagram evaluates to
\begin{align}
\mathcal{M}^{(n)}_\lambda &= \int dz \sqrt{g} \prod_{i=1}^n K(z,k_i)=\ell^{d+1}e^{k_{\Sigma_n} z_c} z_c^{\frac{(d+1) \theta}{d-1}-d} E_{(\nu)}(k_{\Sigma_n} z_c)\,,\\
\textrm{if}\quad n&>\f{2(d^2-\theta-d(1+\theta))}{(d-1)(d-1-\theta)}\,;\quad k_{\Sigma_n} = \sum_{i=1}^n k_i\,,\quad \nu = -\frac{(d (n-2)-n-2) (d-\theta-1)}{2 (d-1)},
\end{align}
for the exponential integral function $E_{(\nu)}(z) = \int_1^\infty dt e^{-zt}/t^\nu$. Expanding this general expression for small $z_c$, the divergent structure is as follows:
\begin{align}
\mathcal{M}_{\lambda}^{(n)}=\,\ell^{d+1}
\left(z_c^{\f{(d+1)\theta}{d-1}-d}+k_{\Sigma_n}z_c^{1+\f{(d+1)\theta}{d-1}-d}+\dots\right.\\
\left.\qquad +k_{\Sigma_n}^{\nu-1}z_c^{\f{n}{2}(1-d+\theta)}+k_{\Sigma_n}^{\nu}z_c^{1+\f{n}{2}(1-d+\theta)}+\dots\right).
\end{align}
There are various special cases of this structure where the powers begin to coincide at some point in the two series. In these cases there are often cancellations and the series truncates at the first common power in the series. 

\subsection{$L$-Loop diagram}
Our general answer is
\begin{align}
\mathcal{M}_{\lambda}^{(n-2L)} &= \int_{z_c}^{\infty} dz \sqrt{g} \left(\prod_{i=1}^{n-2L} K(z,k_i)\right)\int_0^{\Lambda} \left(\prod_{i=1}^L d^d p_i  G(z,z;p_i)\right)\\
&=\left((d-1)\,\Gamma \left(\f{d}{2}\right)\right)^{-L}\ell^{d+1}\Lambda^{L(d-1)}\pi ^{dL/2}   e^{k_{\Sigma_{n-2L}} z_c} z_c^{\f{(d+1) \theta}{d-1}-d} E_{(\nu)}(k_{\Sigma_{n-2L}} z_c)\,,\\
&\nu=\frac{(d-\theta-1) (d (2 L-n+2)-2 L+n+2)}{2 (d-1)}\,.
\end{align}
In doing this calculation, we first do the momentum integral of the bulk-to-bulk propagator $G$ and keep only the large-momentum divergence  $\Lambda^{L(d-1)}$. This piece is independent of $\ell$ and $z_c$. As a result, the answer above, and therefore the small $z_c$ expansion, is the same as in the tree-level diagram, except with the substitution $n \rightarrow n-2L$.

\section{``Conformally massive" scalar}
\label{app:confmass}
It is not possible to solve the wave equation for a massive scalar in FRW in closed form, so we cannot analyze the divergences for that case. However, we can instead consider modifying the coefficient of the nonminimal coupling $R\phi^2$ to be an arbitrary number which we will suggestively call $m^2$ even though it is dimensionless. In this case we can solve the wave equation:
\begin{gather*}
\left(\f{1}{\sqrt{-g}}\p_\mu \left(\sqrt{-g}g^{\mu\nu}\p_\nu\right) + m^2 R\right) \phi = 0,\\
\implies \phi(\eta)=\eta ^{\frac{d-\h}{2}} \left(c_1 J_{\nu}(-i k \eta )+c_2 Y_{\nu}(-i k \eta )\right), \\
 \nu^2=\left(\left(1-4 m^2\right) d^3+\left(\left(8 m^2-2\right) \h-1\right) d^2+\left(\h (\h+2)-4 m^2 \left(\h^2-1\right)\right) d-\h^2\right)(4 (d-1))^{-1}.
\end{gather*}
Notice that this solution looks like that of a massive scalar in de Sitter. We can make the connection a little more precise. For massive scalars in de Sitter we have 
\begin{align}
\phi(\eta)=\eta^{d/2}\left(c_1 J_{\frac{1}{2} \sqrt{d^2+4 \tilde{m}^2 \ell ^2}}(-i k \eta )+c_2 Y_{\frac{1}{2} \sqrt{d^2+4 \tilde{m}^2 \ell ^2}}(-i k \eta )\right)
\end{align}
This means that for 
\begin{align}
m^2=-\frac{(-1 + d) ((3 d - \h) (d + \h)  + 16 \tilde{m}^2 \ell^2)}{
 4 d (-1 + (d-\h^2)^2)}
 \end{align}
 the two solutions are conformally related by a factor of $\eta^{-\h/2}$, just as in the case of a massless scalar.  In other words, for every value of the mass of the dS scalar, there exists a value of the ``conformal mass" of the FRW scalar for which the two solutions are conformally related.

 \small
\bibliographystyle{JHEP}
\bibliography{FRWbib}

\end{document}